\documentclass[12pt]{article}

\input epsf
\def\beq{\begin{eqnarray}}
\def\eeq{\end{eqnarray}}
\def\m{M_*}
\def\mpl{M_{\rm Pl}}

\def\L*{{\cal L}_*}
\def\lsim{\mathrel{\rlap{\lower3pt\hbox{\hskip0pt$\sim$}}
     \raise1pt\hbox{$<$}}}         
\def\gsim{\mathrel{\rlap{\lower4pt\hbox{\hskip1pt$\sim$}}
     \raise1pt\hbox{$>$}}}         

\usepackage{amsmath}
\usepackage{amsfonts}

\begin{document}

\begin{titlepage}

\begin{flushright}
{NYU-TH-06/12/10}
\end{flushright}
\vskip 0.9cm

\centerline{\Large \bf A Model for Cosmic Self-Acceleration}

\vskip 0.7cm
\centerline{\large Gregory Gabadadze}
\vskip 0.3cm
\centerline{\em Center for Cosmology and Particle Physics}
\centerline{\em Department of Physics, New York University, New York, 
NY, 10003, USA}

\vskip 1.9cm

\begin{abstract}

We discuss a model which gives rise to cosmic 
self-acceleration due to modified gravity.
Improvements introduced by this approach
are the following: In the 
coordinate system commonly used, the metric does not grow in 
the bulk, and  no negative mass states
are expected to appear. The spectrum of small perturbations 
contains a  localized massless tensor mode, but does not admit 
dangerous localized massive gravitons.  All the massive spin-2 modes are 
continuum states.  The action of the model, which is an extension of DGP, 
allows to relax the previously known constraint on the bulk 
fundamental scale of gravity. The latter can take any value below the 
4D Planck mass.

\end{abstract}

\vspace{3cm}

\end{titlepage}

\newpage

\section{Introduction}

The DGP model \cite {DGP} admits  the self-accelerated solution
\cite {Cedric} which could be used to describe \cite {DDG} 
the de Sitter (dS) like expansion of the present-day Universe.  

The model exhibits strongly-coupled behavior already 
at the classical level \cite {DDGV}. As a result,
the question of stability of the self-accelerated background w.r.t. 
small perturbations, \cite {Luty} -- \cite{KaloperRuth}, 
becomes difficult \cite {DGI,Dvali}. 

Luckily, certain non-linear solutions have been  found.  
This is a case for a Schwarzschild-like solution, 
for which the 4D metric was  exactly obtained \cite {GI}, 
and for the Domain Wall solution for  which the full 5D 
metric was found in Ref. \cite {DGPR}.  In both cases, 
the mass (tension) of the solution gets screened by gravitational 
effects, and  these sources on the {self-accelerated background} 
look as if they had a {\it negative} net 5D mass 
(tension)\footnote{In contrast with this, screening of 
the similar sources on the conventional branch of DGP,  leaves them with  
positive 5D mass (tension) \cite {GI,DGPR}.}. 
This suggests that the self-accelerated  background  
may not be problem-free in the full non-linear theory. 

What is a root-cause of this behavior?  
The self-accelerated solution exists  only for a certain choice of 
the sign of the extrinsic curvature, 
and this choice is such that it requires a growing metric
in the bulk. For instance, in a simplest spatially-flat 
case, the full 5D metric of the self-accelerated solution takes 
the form \cite {Cedric}:
\beq
ds^2 = \left (1 +  H|y|\right )^2 \left \{ - dt^2 + e^{2Ht}
d{\vec x}^2 \right \}   
+  dy^2\,,
\label{metricSA}
\eeq
where $y$ is the 5th coordinate  and $H$ denotes the dS expansion 
rate of the 4D worldvolume (the latter is labeled here 
by Cartesian coordinates $(t, {\vec x})$, and we use the 5D coordinate 
system in which the brane is located at $y=0$.). 
The unusual feature of the above  metric is that 
it grows in the bulk,  even though the 
worldvolume metric is that of dS space. A linearly growing metric,
similar to (\ref {metricSA}), would have been produced by a {\it negative}
tension 3-brane\footnote{This could be in, e.g.,  
5D Minkowski or Anti de Sitter (AdS) space-time.}. 
The  growing metric (\ref {metricSA}) imprints 
its ``negative'' effects on the brane worldvolume through the 
extrinsic curvature, giving rise to the 
solutions mentioned in the previous paragraph.

The aim of the present work is to modify the DGP equations
in such a way that the new system  admits a background 
that is equivalent to the self-accelerated solution on the 4D brane, 
however, differs from it in the bulk.  

The new solution that we will discuss takes the form:
\beq
ds^2 = \left (1 -  H|y|\right )^2 \left \{ - dt^2 + e^{2Ht} d{\vec x}^2 
\right \} +  dy^2\,.
\label{metricFSA}
\eeq
In order to obtain this solution one  needs
to flip the sign in front of  the extrinsic curvature term 
in one of the DGP equations, keeping the rest of the equations 
intact.  In section 4 we will discuss
how such equations can be obtained by modifying  
the DGP action. The latter will have an additional benefit:
the constraint \cite {DGP} on the bulk fundamental scale of gravity
will be relaxed. This scale,  in the present context, 
can take any value below the 4D Planck mass. 
 
The solution (\ref {metricFSA}) is formally identical 
to that for a  3-brane endowed with a positive 4D cosmological constant 
(brane tension) which is  embedded in 5D empty space in 5D GR \cite {KLinde}.
However, unlike the latter, the worldvolume expansion 
in the present case (\ref {metricFSA})  is due to modified gravity, 
while the 4D cosmological constant is set to zero. 
This difference is what is responsible 
for the modified Friedmann equation, and distinct 
cosmological evolution on the self-accelerated background.

The spectrum of small perturbations on (\ref {metricFSA}) 
will differ from that of the 3-brane with tension, as well as from the 
spectrum on (\ref {metricSA}).  As we will show, the spectrum 
contains  one {\it massless} spin-2 localized mode, 
the coupling of which  on the brane is $\sim 1/\mpl^2$, and a continuum 
of massive  spin-2 modes with $m^2\ge 9H^2/4$, with   
suppressed couplings on the brane, due to the 
4D Einstein-Hilbert (EH) term.

The bulk space in  both (\ref {metricSA}) and (\ref {metricFSA})
is locally equivalent to 5D Minkowski space. In the chosen coordinate system 
the solution (\ref {metricFSA}) encounters the Rindler horizon at 
$|y|=H^{-1}$. However, an  analytic 
continuation beyond this point  can be performed 
by employing new coordinates (see, e.g., \cite {Wald}). In that 
coordinate system the brane (with closed spatial sections) 
can be regarded as a  4D dS bubble that is  first contracting 
and  then re-expanding in 5D Minkowski space.

\section{Flipped equation}

We  briefly review below the DGP equations and self-accelerated solution 
(\ref {metricSA}), and show how modifying one of those equations can give 
rise to the new solution (\ref {metricFSA}).
   
Consider  the 4D DGP equation written at $y=0^+$ (we use conventions 
of \cite {Wald}):
\beq
G_{\mu\nu} - {m_c} (K_{\mu\nu} - g_{\mu\nu} K)
= 8\pi G_NT_{\mu\nu}(x)\,.
\label{junction}
\eeq
Here, $G_{\mu\nu}=R_{\mu\nu} -g_{\mu\nu}R/2$, is the 4D Einstein 
tensor for the metric 
$g_{\mu\nu}(x,y)$;  $K=g^{\mu \nu}K_{\mu \nu}$, is the trace of the 
extrinsic curvature tensor 
\beq
K_{\mu \nu}\,=\,{1\over 2N}\,\left (\partial_y g_{\mu \nu} -\nabla_\mu 
N_\nu
-\nabla_\nu N_\mu \right )\,,
\label{K}
\eeq
and $\nabla_\mu $ is a 4D covariant derivative w.r.t. the metric 
$g_{\mu \nu}(x,y)$. We introduced the {\it lapse} scalar field  $N$, and 
the {\it shift} vector field $N_\mu $ \cite {ADM}:
\beq
g_{\mu 5}\,\equiv \, N_\mu =g_{\mu\nu}N^\nu\,,~~~g_{55}\,\equiv \,N^2\,+\,
g_{\mu\nu}\,N^\mu \,N^\nu\,. 
\label{adm}
\eeq 
Furthermore, the  $\{\mu\nu \}$ equation in the bulk, and 
the $\{\mu 5\}$ and  $\{5 5\}$ equations read respectively 
\beq
G^{(5)}_{\mu\nu} =0\, ~~~{\rm for}~~~y\neq 0\,, 
\label{bulkmunu} \\
\nabla^\mu K_{\mu\nu} = \nabla_\nu K\,,
\label{mu5new} \\
R= K^2 -K_{\mu\nu}K^{\mu\nu}\,.
\label{55} 
\eeq
Here $G^{(5)}_{\mu\nu}$  denotes the 5D Einstein tensor
for the 5D metric $g_{AB}(x,y)$ (A,B=0,1,2,3,{\it 5}), and 
$g_{\mu\nu}(x,y)$  is its 4D part.
Note that the $\{\mu 5\}$ and  $\{5 5\}$ equations,
(\ref {mu5new}) and (\ref {55}),  should be satisfied in 
the bulk, $y\neq 0$ , as well as  on the brane, $y=0$.  

\vspace{0.2in}

Let us turn to cosmological solutions. The metric is parametrized as follows:
\beq
ds^2 = - P^2(t,y) dt^2 + Q^2(t,y) \gamma_{ij} dx^i dx^j + 
\Sigma^2 (t,y) dy^2\,.
\label{metric}
\eeq
The self-accelerated solution reads \cite {Cedric}:
\beq
P(t,y) = 1+|y| { {\ddot a}\over \sqrt {{\dot a}^2 +k} }, ~~
Q(t,y) = a(t) + |y| {\sqrt {{\dot a}^2+k} },~~~
\Sigma (t,y) =1\,.
\label{sas}
\eeq
Here we included a nonzero spatial curvature $k$. 
With this Ansatz, the Friedman equation on the brane 
follows from (\ref {junction}), and can be expressed  
in terms of the 4D Hubble parameter $H\equiv {\dot a}/a$.
For a simplest case of $k=0$ it reads \cite {Cedric}
\beq
H^2 -m_c|H| = 0\,.
\label{Friedmann}
\eeq
This admits  a dS  solution with $H=m_c$. We emphasize that 
the minus sign between the two terms in (\ref {Friedmann}) 
is guaranteed by the choice of the positive sign in front of the 
terms in  (\ref {sas}) that are proportional to $|y|$. 
If we were to  choose the latter signs to be negative, 
we would have obtained the Friedman equation
$H^2+m_c |H|=0$, which does not admit the dS  solution.

On the other hand, as was discussed in section 1, 
it is exactly the positive  sign in front of the 
$|y|$ terms in (\ref {sas})  that gives rise to the  negative mass 
difficulties. Hence, the question: can we avoid these problems 
while still retaining good properties of the self-accelerated 
solution?  

\vspace{0.1in}

The proposal is to flip the sign in front of the second term on 
the l.h.s. of (\ref {junction}). In other words, we introduce 
a new set of equations in which  (\ref {junction}) 
is replaced by:
\beq
G_{\mu\nu}\, + \, {m_c} (K_{\mu\nu} - g_{\mu\nu} K)
= 8\pi G_NT_{\mu\nu}(x)\,,
\label{fjunction}
\eeq
while all the other equations (\ref {bulkmunu}),(\ref {mu5new}) and 
(\ref {55}) remain intact. The action functional that 
gives rise to this new set of equations will be discussed in the next 
section. Here, for the purposes of finding a classical solution 
and perturbations about it,  it is enough to focus  
on the equations of motion.

The metric for the self-accelerated solution of the new 
system of equations (\ref {fjunction}, \ref {bulkmunu}--\ref {55})
reads:
\beq
P(t,y) = 1 - |y| {{\ddot a}\over \sqrt {{\dot a}^2+k}   },
~~~ Q(t,y) = a(t) - |y| \sqrt {{\dot a}^2+k },
~~~\Sigma (t,y) =1\,,
\label{fsas}
\eeq
where we have chosen a negative sign in front of the terms proportional 
to $|y|$. Let us now see how the change of the positive signs 
in the metric (\ref {sas}) into the negative signs in 
(\ref {fsas}) changes the value of the 
extrinsic curvature evaluated at $y=0^+$.
On the solution (\ref {fsas}), $N_\mu=0$, $N=1$, and 
$K_{\mu\nu} = \partial_y g_{\mu\nu}/2$.  Hence, at $y=0^+$ 
the components of the extrinsic curvature tensor evaluated on 
the solution (\ref {fsas}) equal to {\it minus}  the 
corresponding components evaluated on (\ref {sas}).  
Therefore, substitution of  (\ref {sas})
into  (\ref {junction})  would give the same 
equation as the substitution of (\ref {fsas}) into (\ref {fjunction}).
The corresponding Friedmann equation on the {\it empty} brane,
which now follows from  (\ref {fjunction}) instead of (\ref {junction}), 
reads:
\beq
H^2 + {k\over a^2}= m_c \sqrt {H^2 + {k\over a^2}}\,. 
\label{Fr}
\eeq
For $k=0$ this coincides with (\ref {Friedmann})
and gives the spatially-flat  dS solution with $H=m_c$ (\ref {metricFSA}). 
For general $k$  the solutions are: 
\beq
ds^2 = \left (1- H|y|\right )^2 \left \{ - dt^2 + a^2(t)
(d\chi^2 + S^2_k(\chi) d \Omega^2)  \right \}   +  dy^2\,,
\label{metricSol}
\eeq
where $H=m_c$ and  $k=-1,0,1$ corresponds to the open, flat and closed spatial 
slicing of 4D dS space, for which  $S_k(\chi)= {\rm sinh} \chi,  
\chi, {\rm sin} \chi$, respectively\footnote{There are 
two other solutions to (\ref {Fr}). For $k=0$ one finds the  $H=0$
flat solution. For $k=-1$ one finds the Milne solution $a(t)=t$.}.

The solution  (\ref {metricSol})  should satisfy all the bulk 
equations (\ref {bulkmunu} -- \ref {55}), since in the bulk it is 
locally equivalent to Minkowski space. We  checked by direct 
substitution that (\ref {metricSol}) solves Eqs. 
(\ref {bulkmunu} -- \ref {55}) too. 

It is also straightforward to introduce matter/radiation on the brane.
Following \cite {Cedric} we obtain  the Friedmann equation
\beq
H^2+{k\over a^2} = \left (\sqrt{{8\pi G_N\over 3}\rho +{m_c^2\over 4}} + 
{m_c \over 2}\right )^2\,, 
\label{matter}
\eeq
which should be amended by  the conventional conservation equation 
for the fluid of density  $\rho$  and pressure $p$: ${\dot \rho} + 
3H(\rho +p)=0$. The latter being a result  of the 
matter stress-tensor conservation $\nabla^\mu T_{\mu\nu}=0$, which 
can be verified, e.g., by taking a covariant derivative 
of both sides of (\ref {fjunction}) and using  (\ref {mu5new}).

\section{Perturbations}

In this section we study the spectrum of small perturbations
of the theory (\ref {fjunction}, \ref {bulkmunu} -- \ref {55}).
Following \cite {DGI} we consider the metric perturbations of the form:
\beq
ds^2 = \left ( {\bar g}_{\mu\nu}(x,y) + \delta g_{\mu\nu}(x,y) \right )
dx^\mu dx^\nu + 2 \delta g_{\mu 5}dx^\mu dy+ 
\left (1+ \delta g_{55}(x,y)   \right )dy^2\,. 
\label{metricg}
\eeq
The background metric will be distinguished by the bar
$ {\bar g}_{\mu\nu}(x,y) =A^2(y) \gamma_{\mu\nu}(x)$, where 
$A\equiv (1-H|y|)$, and $\gamma_{\mu\nu}(x)$ is a metric tensor
for 4D dS space-time.
 
Let us consider  small perturbations about an {\it empty} (i.e., 
with $T_{\mu\nu} =0$)  self-accelerated background (\ref {metricSol})
for $k=0$. The following expansion of the metric and 
extrinsic curvature tensors will be used:
\beq
g_{\mu\nu} = {\bar g}_{\mu\nu} +\delta g_{\mu\nu}\,,~~~K_{\mu\nu}
= {\bar K}_{\mu\nu} + \delta K_{\mu\nu} \,, \\
{\bar K}_{\mu\nu} =n {\bar g}_{\mu\nu} \,,~~~ {\bar K} = 
{\bar g}^{\mu\nu} {\bar K}_{\mu\nu} =4 n\,~~~n\equiv {\partial_y A\over A}\,,
\label{fluct}
\eeq
where
\beq
K = g^{\mu\nu}K_{\mu\nu} \equiv {\bar K} + \delta K\,, 
~~~ \delta K = {\bar g}^{\mu\nu}\delta K_{\mu\nu}
-  n {\bar g}^{\mu\nu} \delta g_{\mu\nu}\,. 
\label{Knew}
\eeq 
It is straightforward to check that the perturbations of the 
off-diagonal equation (\ref {mu5new}) satisfy  the following relation
\beq
\nabla^\mu \left (  \delta K_{\mu\nu} - n \delta g_{\mu\nu}\right )=
 \nabla_\nu \left ({\bar g}^{\rho\sigma} \delta K_{\rho\sigma} -
n {\bar g}^{\rho\sigma} \delta g_{\rho\sigma} \right )\,,
\label{mu5pert}
\eeq
where in this section $\nabla $ denotes  a 4D covariant derivative
for ${\bar g}_{\mu\nu}$. 

Furthermore, small perturbations of the $\{55\}$ 
equation (\ref {55}) yield
\beq
\delta R = 6 n ( {\bar g}^{\mu\nu}\delta K_{\mu\nu}
-n {\bar g}^{\mu\nu} \delta g_{\mu\nu})\,,
\label{55pert}
\eeq
while from the source-free junction condition (\ref {fjunction}) 
we obtain
\beq
\delta R|_{0^+} = - 3 m_c ( {\bar g}^{\mu\nu}\delta K_{\mu\nu}
-n {\bar g}^{\mu\nu} \delta g_{\mu\nu})|_{0+} \,. 
\label{junctpert}
\eeq
Since $m_c=H$, equations (\ref {55pert}) and (\ref {junctpert})
are in contradiction with each other unless the r.h.s. 
of (\ref {junctpert}) is zero. Requiring also a 
continuity of $\delta R$  we  find  
that the r.h.s. of  (\ref {55pert}) should 
be zero for arbitrary $y$:
\beq
{\bar g}^{\mu\nu}\delta K_{\mu\nu} -n  {\bar g}^{\mu\nu} 
\delta g_{\mu\nu}=0\,. 
\label{sourcefree}
\eeq
Finally, let us look at small perturbations of the bulk $\{\mu\nu \}$ 
equation.  As long as $T_{\mu\nu}=0$ 
we can use  the Gaussian  normal coordinates and simultaneously 
choose the  following ``gauge''\footnote{In general, this is not an acceptable 
gauge fixing condition  when the brane is held straight and 
the metric is coupled to a non-conformal source. In the latter 
case  one  needs to introduce the brane bending mode \cite {GT}. 
We'll briefly comment on this below.}
\beq
\nabla^\mu \delta g_{\mu\nu} =0\,,~~~ {\bar g}^{\mu\nu}\delta 
g_{\mu\nu} =0\,,~~~ \delta g_{55}= \delta g_{\mu5} =0\,.
\label{ttgauge}
\eeq
Under these conditions the bulk $\{\mu\nu \}$ equation takes 
the form:
\beq
\partial_y^2 \delta g_{\mu\nu} + {1\over A^2} 
\left (\square_4 -4H^2 \right )\delta g_{\mu\nu} =0\,,
\label{bulkpert}
\eeq
while the junction condition reads as follows:
\beq
-{1\over 2} \left (\square_4 -4H^2 - H\partial_y \right )\delta
g_{\mu\nu}|_{0^+}=0\,.
\label{junction00}
\eeq
Having these equations specified, we can calculate the spectrum of
Kaluza-Klein (KK)  modes. It is convenient to 
use the following decomposition:
\beq
\delta g_{\mu\nu} (x,y) 
\equiv \int h_{\mu\nu}^{(m)}(x) {\tilde f}_m(y) 
dm\,,~~~~~( \square_4 -2H^2 )h_{\mu\nu}^{(m)}(x)  =m^2 
h_{\mu\nu}^{(m)}(x)\,.
\label{KK}
\eeq
By introducing  a  new coordinate $z$ and a  function $u_m$
\beq
dz \equiv {dy \over A(y)}\,, ~~~~ {\tilde f}_m \equiv A^{1/2} u_m \,, 
\label{new}
\eeq
we turn  (\ref {bulkpert}) into a 
Schr\"odinger-like equation with a 
boundary condition set by (\ref {junction00}).
The former reads 
\beq
-{d^2u_m  \over dz^2}  + \left ({9H^2\over 4} -m^2  \right )u_m =0\,,
\label{sbulku}
\eeq
while the boundary  condition  takes the form:
\beq
[\partial_z u_m(z)]^{0^+}_{0^-} = -\left 
(3H - 2 {m^2\over H} \right )u_m(0)\,.
\label{sjunctionu}
\eeq
The spectrum of (\ref{sbulku}, \ref {sjunctionu}) can be 
computed. The condition for the modes to be 
localized on the brane translates into the requirement, 
$\int_{-\infty}^{+\infty} dz u^2_m(z)<\infty$.

The above system admits a localized zero-mode solution.
For  $m=0$,  we obtain $u=c \,{\rm exp}(-3H|z|/2)$ 
($c$ being an  constant). The zero-mode is normalizable. 

Furthermore, one can check that there are no other normalizable modes 
satisfying both (\ref{sbulku}) and (\ref {sjunctionu}).
However, there are an infinite number of continuum states with  
masses $m_{KK}^2\ge 9H^2/4$. These are plane-wave normalizable
in the $z$ coordinate, and their wave-functions are suppressed on the 
brane because of the 4D EH term \cite {DGKN} (a similar spectrum for a scalar 
was discussed in \cite {Oriol}).

Let us compare the above-obtained  spectrum to  that  
on the selfaccelerated background (\ref {metricSA}), found in 
\cite {Koyama}. The two spectra are similar to each other 
for $m_{KK}^2\ge 9H^2/4$, as both of them consist of massive KK modes
with suppressed couplings on the brane.  However, the 
localized modes differ. On the background (\ref {metricSA})
one finds a localized {\it massive} mode with $m_*^2 =2H^2$, while the 
localized mode on (\ref {metricFSA}) is {\it massless}, $m_0=0$. 
In both massive and massless cases there is  an enhanced 
``gauge symmetry'' for the localized modes:  
in  the massive case  the symmetry transformation is \cite {Deser}
\beq
\delta h^*_{\mu\nu}(x) = 
(\nabla_\mu\nabla_\nu +H^2 \gamma_{\mu\nu})\alpha(x)\,,
\label{enh}
\eeq
(where $\alpha$ is a gauge transformation parameter)
while for the massless case this is just a conventional 4D
reparametrization  invariance  $\delta h^0_{\mu\nu}(x) = 
\nabla_\mu \zeta_\nu(x) + \nabla_\nu \zeta_\mu(x) $. 
The symmetry transformation (\ref {enh}) is what allows one 
to gauge away (in the quadratic approximation and on an empty brane) 
the dynamics of the brane bending mode, converting the latter into 
a Lagrange multiplier  field \cite {DGI}. However, the symmetry (\ref {enh}) 
is broken once  non-conformal sources 
are switched on. This leads to: (a) propagation of the 
bending mode with potential  ghost-like instability; 
(b) breakdown of perturbation  theory making the linearized 
calculations unreliable \cite {DGI,Dvali}.

Things are  different in the present case, since  
we have a localized massless mode. The gauge 
symmetry for it, $\delta h^0_{\mu\nu}(x) = 
\nabla_\mu \zeta_\nu + \nabla_\nu \zeta_\mu $, is preserved 
when the sources are switched on. Then, one 
should either take into account the 
brane bending mode, or perform calculations with other 
acceptable gauge conditions, such as e.g., the 5D 
harmonic gauge. The expectation is that this
won't introduce an additional on-shell degree of freedom.
This will be discussed in detail elsewhere.

\section{Modified action}

In this section we  will discuss how one should modify the DGP action
in order to flip  the sign in front of the extrinsic curvature 
terms in the junction condition, i.e., to replace (\ref {junction}) 
by (\ref {fjunction}). Naively, this can be done by flipping the 
sign of the 5D Einstein-Hilbert term in the DGP action. However, 
this would lead to a bulk graviton with 
a wrong-sign  kinetic term.  This is not a  good course of 
action.

Instead, we will introduce an 
additional 5D Einstein-Hilbert term on the brane, with a small 
{\it negative}  coefficient.  This is acceptable as long as there 
is a  4D Einstein-Hilbert term on the brane with a large positive 
coefficient. We will show that the effect of the new term is twofold: 
(I) If allows to 
flip the desired sign in the junction condition; (II) 
It relaxes the constraint on the bulk fundamental scale $\m$;
the latter can now take an arbitrary value below $\mpl$, as long as 
the coefficient of the brane localized 5D Einstein-Hilbert term 
is tuned appropriately.

We will discuss these properties first on a simple scalar field example.
Then we elevate the construction to gravity.

\subsection{Scalar example}

Let us start with  the scalar analog of the DGP action \cite {DGP}
\beq
-{\mpl^2 \over 2} \int d^4x  
(\partial_\mu \phi(x,0))^2 - {\m^3 \over 2}
\int d^4x dy (\partial_A \phi(x,y))^2\,,
\label{scalar1}
\eeq
where the dimensionless scalar field $\phi$ is 
to mimic a graviton. We impose ${\bf Z_2}$ symmetry across
the $y=0$ boundary,  and add to the above action 
the coupling of $\phi$ to  a source $J$, also localized on the brane,    
$\int d^4x dy \delta (y) J\phi$. To get the junction condition, 
we integrate the equation of motion obtained  from  
(\ref {scalar1}) w.r.t. $y$, from $0^{-}$ to $0^+$.
The resulting equation written at $y=0^+$ reads:
\beq
- \partial^2_\mu \phi|_{y=0}  - m_c \partial_y\phi|_{y=0^+}= J/\mpl^2\,.  
\label{junctionscalar}
\eeq
Here we introduced the definition $m_c = 2\m^3/\mpl^2$, and for 
comparison with gravity, flipped the overall sign of the equation.

Our goal is to modify the action (\ref {scalar1})
so that the resulting junction condition has 
an opposite sign in front of the second term on 
the l.h.s of (\ref {junctionscalar}).

For this we introduce an additional term into 
the action (\ref {scalar1}). This is just a 5D kinetic term
peaked  on the brane. To make things tractable, we 
smear the brane, that is, instead of the Dirac function $\delta (y)$,
we use its regularized version  $\delta (y)\to {\bar \delta}(y) 
\equiv \pi^{-1} \epsilon/(y^2 +\epsilon^2)$, with $\epsilon \to 0$.
The term that we'll be adding to (\ref {scalar1}) then reads:
\beq
{{M}^2\over 2}  \int dx dy {\bar \delta}(y) (\partial_A \phi(x,y))^2\,.
\label{scalaradd}
\eeq
With this term included the variation of the action $\delta S=0$ 
with the appropriate boundary conditions gives:
\beq
- (\mpl^2 -M^2) {\bar \delta}(y)  \partial^2_\mu \phi  -\m^3 \partial^2_\mu
\phi - \partial_y\left ((\m^3 -M^2 {\bar \delta}(y)) 
\partial_y \phi \right ) = 
J{\bar \delta}(y) \,.
\label{fjunctionscalar0}
\eeq
Next we take  the integral of both sides of this equation w.r.t. $y$ from
$-\epsilon $ to $+\epsilon $, and then turn to the limit 
\beq
M\to 0,~~~\epsilon \to 0,~~~ M^2/\epsilon \sim 
M^2 {\bar \delta} (0)\equiv {\bar M}^3> {\m^3}\,, 
\label{limit}
\eeq
where we keep ${\bar M}$ fixed, and its value  
bigger than $\m$. The resulting equation reads:
\beq
-  \partial^2_\mu \phi
+{2 ({\bar M}^3-  \m^3) \over \mpl^2}  \partial_y \phi = J/\mpl^2\,.
\label{fjunctionscalar1}
\eeq
Finally, introducing
\beq
m_c \equiv  {2 ( { {\bar M}^3 - \m^3 }) \over \mpl^2}\,,
\label{newmc}
\eeq
where the positive numerical value of $m_c$ will be tuned 
to the Hubble scale today  $m_c\sim H_0 \sim 10^{-42}$ GeV,
we get the desired junction conditions
\beq
- \partial^2_\mu \phi  + m_c \partial_y\phi = J/\mpl^2\,.  
\label{fjunctionscalar}
\eeq
The second term on the l.h.s. of (\ref {fjunctionscalar})
has a sign opposite to the one of the analogous term 
in (\ref {junctionscalar}). This accomplishes our goal
\footnote {The Minkowski space propagator following 
from  (\ref {fjunctionscalar}), would have a resonance-like 
pole on the first Riemann sheet, signaling very slow 
tachyonic instability \cite {GG}. 
The gravitational counterpart of this theory, also admits  
Minkowski space as a solution, in addition to the dS solution 
we are interested in. Perturbations about
the former  will have a  similar pole on the first Riemann
sheet, showing instability of the Minkowski solution.  This 
instability should be more severe then in the scalar case, 
as the  tachyon-like pole here would appear for a tensor state
(i.e., the longitudinal scalar mode would acquire a ghost-like 
kinetic term). This is encouraging, since our eventual goal is 
to have only the stable dS background. In the 
latter case the  pole structure of the propagator is modified 
due to  the background. The kinetic term of the longitudinal mode
should get cured by the dS curvature effects.}.

Two comments are in order here.  First, 
the wrong-sign kinetic term (\ref {scalaradd}) that is peaked only 
on the brane is dominated  by 
the large positive 4D kinetic term in (\ref {scalar1}), 
proportional to $\mpl^2 > M^2$. Second, the number of adjusted parameters 
here is the same as in DGP:
In (\ref {scalar1}) one should tune the value of $\m$ such that the 
ratio $2\m^3/\mpl^2$ is of order $H_0$. On the other hand, 
in the action with the additional  term (\ref {scalaradd}) the value of 
$\m<\mpl$ can be arbitrary, however, for a given value of 
$\m$ one should tune the value of  ${\bar M}$ so that 
(\ref {newmc}) is of order $H_0$.
In the case of gravity, to which we'll turn in the next subsection,
this has an additional benefit of relaxing the constraint 
on the bulk  scale of gravity, $\m$, which now can  be an arbitrary 
scale somewhat lower than $\mpl$.

\subsection{Gravity}

The DGP gravitational action is  \cite {DGP}
\beq
S\,=\,{\mpl^2 \over 2} \,\int\,d^4x\, \sqrt{g_4}\,R(g_4)\,+
{\m^{3} \over 2}\,\int \,d^4x\,dy\,\sqrt{g_5}
\,{\cal R}(g_5)\,,
\label{1}
\eeq
where $g_{4\mu\nu}\equiv g_{\mu\nu}(x,0)$, and 
 $g_5$ refers to the full 5D metric; $R$ and ${\cal R}$ are 
the four-dimensional
and five-dimensional Ricci scalars respectively, and
$\m$, as before,  stands for the fundamental gravitational scale
of the bulk theory.  The brane  is located  at $y=0$ and ${\bf Z}_2$
symmetry across the brane is imposed. The 
boundary Gibbons-Hawking term should be taken into account  
to warrant the correct Einstein equations in the bulk.
The matter fields, that are also omitted here for simplicity, 
are assumed to be localized on the brane.

The above action gives rise to the equations  of motion
(\ref {junction},\ref {bulkmunu}-\ref {55}). 
As we discussed previously, we'd like our 
equations to be (\ref {fjunction}, \ref {bulkmunu}-\ref {55}) 
instead. To reach this goal, we proceed as in the scalar case 
discussed in the previous subsection.

We smooth out the brane by  replacing 
$\delta (y) \to {\bar \delta}(y)$,   and  add the following 
boundary (worldvolume) term to the action:
\beq
-{M^2\over 2} \int \,d^4x\,dy\,{\bar \delta} (y) \sqrt{g_5}{\cal R}(g_5)=
-{M^2\over 2} \int \,d^4x\,dy\,{\bar \delta} (y) \sqrt{g}\, N 
\left (R+K^2-K_{\mu\nu}^2\right )\,,
\label{add}
\eeq
where the r.h.s. of (\ref {add}) is obtained by using the standard 
ADM decomposition. The total action in the ADM formalism reads:
\beq
S_{\rm mod}\,=\,{\mpl^2 \over 2} \,\int\,d^4x\, 
\sqrt{g_4}\,R(g_4)\,+ {\m^{3} \over 2} \int \,d^4x\,dy\,\sqrt{g}
N(R+ K^2 -K_{\mu\nu}^2)\, \nonumber \\
- {M^{2}\over 2}  \int \,d^4x\,dy\, {\bar \delta} (y)   \sqrt{g} 
N \left (R+ K^2 -K_{\mu\nu}^2 \right )\,,
\label{fdgp}
\eeq
where it is implied that the 4D EH term is also 
smeared over the same scale as the 5D term
\footnote{For the regularization of 4D and 5D EH terms, see,
\cite {DHGS} and  \cite {PR}, respectively.}. 
The equations of motion are 
straightforward to derive from (\ref {fdgp}).
The $\{\mu\nu\}$ equation in the bulk, and $\{\mu 5\}$ and $\{5 5\}$ equations
read  as follows:
\beq
G^{(5)}_{\mu\nu} =0\, ~~~{\rm for}~~~|y|> \epsilon \,, 
\label{fbulkmunu} \\
(\m^3 - M^2 {\bar \delta}(y))
\left (  \nabla^\mu K_{\mu\nu}  - \nabla_\nu K \right ) =0\,,
\label{fmu5new} \\
(\m^3- M^2 {\bar \delta}(y))\left (R - K^2 +K_{\mu\nu}K^{\mu\nu}\right ) =0\,.
\label{f55} 
\eeq
As in the scalar case, we  will be looking at this theory 
in the  limit (\ref {limit}).
The above equations reduce to (\ref {bulkmunu}-\ref {55})
\footnote{Eqs. (\ref {fmu5new}) and (\ref {f55})  also flip the overall 
sign on the brane for our choice of $\bar M$. 
This is important for understanding the  perturbations on 
the self-accelerated background.}.

The Israel condition across the brane gets modified
because of the new term in (\ref {fdgp}).  In the limit 
(\ref {limit}) the junction condition reads:
\beq
G_{\mu\nu} + {2({{\bar M^3} -  \m^3)}\over \mpl^2} 
(K_{\mu\nu} - g_{\mu\nu} K) = 8\pi G_NT_{\mu\nu}(x)\,.
\label{junctionnew}
\eeq
If ${\bar M} =0$, as in (\ref {1}),  we get back the result
(\ref {junction}) with $m_c = 2\m^3/\mpl^2$. However,
${\bar M} $ does not have to be zero. For an 
arbitrary value of $\m$ we tune the value of ${\bar M}$
so that  the crossover scale (\ref {newmc}), 
which  appears in (\ref {junctionnew}),  
is adjusted to the value of the present-day Hubble scale 
$m_c\sim H_0 \sim 10^{-42}$ GeV. Hence, (\ref {junctionnew})
recovers (\ref {fjunction}).

A few important comments:

(I) The number of adjusted parameters here is the same 
as in DGP. In (\ref {1}) one should tune the value of $\m$, so  that the 
ratio $2\m^3/\mpl^2$ is of order $H_0$. This constrains 
$\m \sim 100$ MeV. On the other hand, in (\ref {fdgp})
the value of $\m$ is arbitrary. In order to get the right 
crossover scale  for a given value of $\m$,  one should 
tune the value of  ${\bar M}$ so that  (\ref {newmc}) is of order $H_0$. 
This has an additional benefit that the bulk  scale of gravity,  
$\m$, can  take an  arbitrary value $\m< \mpl$.

(II) The limit (\ref {limit}) is a mathematical 
simplification. In reality the brane will have a 
width $\epsilon$ (in simplest cases $\epsilon \sim \m^{-1}$). 
If a nonzero brane width effects are 
kept track of, then the junction condition (\ref {junctionnew}) will contain
additional terms proportional to $M^2$.  
For instance, among other terms, there'll be ones proportional to
\beq
{M^2\over \mpl^2} \left ( G_{\mu\nu} - {1\over 2} g_{\mu\nu}
(K^2-K_{\mu\nu}^2) - 2 (K_\mu^\alpha K_{\alpha \nu} -K K_{\mu\nu}) 
\right )\,.
\label{terms}
\eeq
Such terms can be neglected as long as $ M\ll \mpl $. 
On the other hand, if  $M \lsim \mpl$
the above terms can introduce modifications in the Friedman 
equation (\ref {Friedmann}), and/or need to rescale the 
overall Newton's constant.

(III)  In the above construction the  function 
$F(y)\equiv (\m^3 - M^2 {\bar \delta} (y))$  asymptotes to
$\m^3$ away from the brane, $|y|\gg \epsilon$, 
while it approaches the negative value, 
$(\m^3 -{\bar M}^3)<0$ when  $y \to 0$. 
Being a continuous function,  $F(y)$  should pass through zero.
On the other hand, $F(y)$ is a function that determines the 
coefficient in front of the 5D EH term in the action (\ref {fdgp}).
This may rise some concerns. Within the brane, i.e., 
for $|y|\lsim \epsilon$ the  5D EH term flips the sign, and the 
concern  may be that this  could  lead to some negative-sign 
kinetic terms.  Precisely when the function $F(y)$ flips 
its sign, a large positive-sign 4D EH term (\ref {fdgp}) becomes 
peaked on the brane. The latter dominates, as long as $\mpl^2 > M^2$.  
Moreover, the kinetic term of the longitudinal modes,
which cannot be helped by the 4D EH term, would become 
normal due to the dS background effects, as the calculations of the 
previous section suggest\footnote{One could  consider 
adding to the DGP action (\ref {1}) only the 
$M^2F(y)( K^2 -K_{\mu\nu}^2)$ term. However, this would also change 
the $\{55\}$ equation in addition to modifying the junction  condition. 
Then, one would need to study the new background solutions.}.  

A related  concern may arise 
in the region where $F(y)$ approaches zero. There, 
the additional helicity states of graviton could become 
infinitely strongly coupled in a {\it naive} 
perturbation theory.  However, as in the case of massive gravity, and 
the DGP model, one should expect that   a smooth transition 
to the $F(y)\to 0$ limit is recovered in  the full nonlinear 
theory \cite {Arkady,DDGV}. By now there are a few examples  and arguments 
supporting  the above expectation, see, e.g., 
\cite {LueString,Andrei,Tanaka}, \cite {GI,DGPR}.

\section{Discussions and outlook}

We discussed a model defined by the equations 
(\ref {fjunction}, \ref {bulkmunu}--\ref {55}), or by the
action (\ref {fdgp}), and show that
it has  the  self-accelerated solution  with decreasing 
metric in the bulk (\ref {metricFSA}). 
The spectrum of small perturbations on the
background (\ref {metricFSA})  is 
different from the spectrum on  (\ref {metricSA}).
The lowest mode is a localized massless spin-2 state.
The low energy theory of this mode exhibits 4D general 
covariance. Most importantly, however, there is no massive localized 
spin-2 state, the one that causes difficulties for the background 
(\ref {metricSA}) when non-conformal sources are switched on it 
\cite {DGI}.   All the massive states in the present case 
appear above the threshold, $m^2\ge 9H^2/4$.

Nonlinear dynamics of this model is expected to 
have features similar to DGP, as well as different ones. 
The differences may arise because of the zero mode. 
On the other hand, the KK gravitons  
form a massive (though light $m \sim H$) 
resonance-like state which should participate, along with the zero mode,
in a one-graviton exchange amplitude. This would lead 
to the vDVZ discontinuity \cite {vDVZ} in the linearized 
theory. However, the continuity should be expected to be 
recovered in the full nonlinear theory \cite {DDGV}, with small, 
but potentially  measurable deviations  from 4D GR 
\cite {DGZ,Lue} (see also, \cite {Iorio1,GI}).

Is the obtained background (\ref {metricFSA}) 
acceptable  from the point of view of
observational cosmology? The time  evolution in (\ref {metricFSA})
is equivalent to that of (\ref {metricSA}). The latter 
has been studied in detail \cite {DDG},\cite {Landau}-\cite {Hu}.  
This evolution is rather 
restrictive.  For instance, the equation of state parameter
today,  $w_0\equiv w(z=0)$, gets related to the matter density $\Omega_M$, as
$w_0 = -1/(1+ \Omega_M )$ (this is for the $k=0$ case \cite {DDG}).   
Fitting the supernova data  with such an  expansion favors 
a low value for $\Omega_M$,  which  is in tension 
with the determination  of the matter density from CMB observables 
\cite {Hu}. It has been suggested in \cite {Hu} to use the open 
($k=-1$) universe to obtain a better fit. Forthcoming measurements 
of the low $z$ supernovae  (which are important as  the calibration 
points) may affect the above considerations.  Moreover, the issues of 
determination of the matter content of the Universe may be influenced by
the non-perturbative phenomena leading to the  mass 
screening \cite {GI,DGPR}. The questions of how much screening
(or anti-screening ) is there, will depend  on a particular model at hand. 
These issues have not been understood yet in the present context.

Finally, an additional benefit of the new term in 
(\ref {fdgp}) is that it allows to relax the constraint on the 
bulk gravity scale. The latter can take an arbitrary value
below $\mpl$. This opens a window for a possible string theory 
realization of this model, or its  $D>5$ counterparts \cite {DG,GS}
(for earlier proposals see \cite {Ignat,Pierre}).

With the new term included, the analog of the 
conventional (Minkowski) solution  of DGP 
could also be studied.  In the latter case, the function 
$F(y)$ should not  flip its sign as it interpolates from its 
bulk value,  $\m^3$, down  to lower {\it positive} value 
on the brane  $\m^3 -{\bar M}^3$ ($F(y)$ could emerge as 
a profile of some 5D scalar field which couples to 5D EH term). 
It would be interesting to  study the nonperturbative solutions, 
similar to those of \cite {GI}, \cite {DGPR}, in this context.

\vspace{0.1in}

\subsection*{Acknowledgments}

The author thanks G. Dvali, O. Pujol\'as and R. Scoccimarro
for useful discussions. The work is supported by NASA grant NNGG05GH34G 
and NSF grant 0403005.


\begin{thebibliography}{99}

\bibitem{DGP}
G.~Dvali, G.~Gabadadze and M.~Porrati,
Phys.\ Lett.\  {\bf B485}, 208 (2000)
[hep-th/0005016].

\bibitem{Cedric}
C.~Deffayet,
Phys.\ Lett.\ B {\bf 502}, 199 (2001)
[arXiv:hep-th/0010186].

\bibitem{DDG}
C.~Deffayet, G.~R.~Dvali and G.~Gabadadze,
Phys.\ Rev.\ D {\bf 65}, 044023 (2002)
[astro-ph/0105068].


\bibitem{DDGV} C.~Deffayet, G.~R.~Dvali, G.~Gabadadze and A.~I.~Vainshtein,
  Phys.\ Rev.\ D {\bf 65}, 044026 (2002)
  [arXiv:hep-th/0106001].


\bibitem{Luty} 
M.~A.~Luty, M.~Porrati and R.~Rattazzi,
JHEP {\bf 0309}, 029 (2003)
[arXiv:hep-th/0303116].


\bibitem{Koyama} 
K.~Koyama,
  Phys.\ Rev.\ D {\bf 72}, 123511 (2005)
  [arXiv:hep-th/0503191].


\bibitem{Gorbunov} D.~Gorbunov, K.~Koyama and S.~Sibiryakov,
  Phys.\ Rev.\ D {\bf 73}, 044016 (2006)
  [arXiv:hep-th/0512097].


\bibitem{KaloperRuth} C.~Charmousis, R.~Gregory, N.~Kaloper and A.~Padilla,
  JHEP {\bf 0610}, 066 (2006)
  [arXiv:hep-th/0604086].

\bibitem{DGI} C.~Deffayet, G.~Gabadadze and A.~Iglesias,
  JCAP {\bf 0608}, 012 (2006)
  [arXiv:hep-th/0607099].

\bibitem{Dvali} G.~Dvali,
  arXiv:hep-th/0610013.



\bibitem{GI}
G.~Gabadadze and A.~Iglesias,
  Phys.\ Rev.\ D {\bf 72}, 084024 (2005)
  [arXiv:hep-th/0407049]; 
  Phys.\ Lett.\ B {\bf 632}, 617 (2006)
  [arXiv:hep-th/0508201].
Phys.\ Lett.\ B {\bf 639}, 88 (2006) [arXiv:hep-th/0603199]. 

\bibitem{DGPR} G.~Dvali, G.~Gabadadze, O.~Pujolas and R.~Rahman,
  arXiv:hep-th/0612016.


\bibitem{KLinde} N.~Kaloper and A.~D.~Linde,
  Phys.\ Rev.\ D {\bf 59}, 101303 (1999)
  [arXiv:hep-th/9811141].

\bibitem{Wald}
  R.~M.~Wald,
  ``General Relativity,'' 1984.

\bibitem{ADM}
  R.~Arnowitt, S.~Deser and C.~W.~Misner,
"Gravitation: an introduction to current research", L. Witten ed. 
(Wiley 1962), pp 227--265;
  arXiv:gr-qc/0405109.


\bibitem{GT} J.~Garriga and T.~Tanaka,
  Phys.\ Rev.\ Lett.\  {\bf 84}, 2778 (2000)
  [arXiv:hep-th/9911055].

\bibitem{DGKN} G.~R.~Dvali, G.~Gabadadze, M.~Kolanovic and F.~Nitti,
  Phys.\ Rev.\ D {\bf 64}, 084004 (2001)
  [arXiv:hep-ph/0102216].


\bibitem{Oriol}  O.~Pujolas,
  JCAP {\bf 0610}, 004 (2006)
  [arXiv:hep-th/0605257].


\bibitem{Deser}S.~Deser and R.~I.~Nepomechie,
  Annals Phys.\  {\bf 154}, 396 (1984). 
S.~Deser and A.~Waldron,
  Phys.\ Rev.\ Lett.\  {\bf 87}, 031601 (2001)
  [arXiv:hep-th/0102166].


\bibitem{GG}
G.~Gabadadze, 
  arXiv:hep-th/0408118; In Ian Kogan Memorial Volume, 
Shifman, M. (ed.) et al. World Scientific, 2004; vol.2, pp 1061-1130. 

\bibitem{DHGS} G.~Dvali, G.~Gabadadze, X.~r.~Hou and E.~Sefusatti,
  Phys.\ Rev.\ D {\bf 67}, 044019 (2003)
  [arXiv:hep-th/0111266].

\bibitem{PR} M.~Porrati and J.~W.~Rombouts,
  Phys.\ Rev.\ D {\bf 69}, 122003 (2004)
  [arXiv:hep-th/0401211].


\bibitem{Arkady} A.~I.~Vainshtein,
  %
  Phys.\ Lett.\ B {\bf 39} (1972) 393.


\bibitem{LueString} A.~Lue,
  Phys.\ Rev.\ D {\bf 66}, 043509 (2002)
  [arXiv:hep-th/0111168].


\bibitem{Andrei} A.~Gruzinov,
  New Astron.\  {\bf 10}, 311 (2005)
  [arXiv:astro-ph/0112246].


\bibitem{Tanaka} T.~Tanaka,
  Phys.\ Rev.\ D {\bf 69}, 024001 (2004)
  [arXiv:gr-qc/0305031].


\bibitem{vDVZ} H.~van Dam and M.~J.~G.~Veltman,
  %
  Nucl.\ Phys.\ B {\bf 22}, 397 (1970); 
V.~I.~Zakharov,
  %
  JETP Lett.\  {\bf 12} (1970) 312
  [Pisma Zh.\ Eksp.\ Teor.\ Fiz.\  {\bf 12} (1970) 447].


\bibitem{DGZ} G.~Dvali, A.~Gruzinov and M.~Zaldarriaga,
Phys.\ Rev.\ D {\bf 68}, 024012 (2003)
[arXiv:hep-ph/0212069].


\bibitem{Lue} 
A.~Lue and G.~Starkman,
Phys.\ Rev.\ D {\bf 67}, 064002 (2003)
[arXiv:astro-ph/0212083].


\bibitem{Iorio1}  L.~Iorio,
  Class.\ Quant.\ Grav.\  {\bf 22}, 5271 (2005)
  [arXiv:gr-qc/0504053]; 
  JCAP {\bf 0509}, 006 (2005)
  [arXiv:gr-qc/0508047];
  JCAP {\bf 0601}, 008 (2006).



\bibitem{Landau} C.~Deffayet, S.~J.~Landau, J.~Raux, 
M.~Zaldarriaga and P.~Astier,
  Phys.\ Rev.\ D {\bf 66}, 024019 (2002)
  [arXiv:astro-ph/0201164].



\bibitem{Lue2} A.~Lue, R.~Scoccimarro and G.~Starkman,
  Phys.\ Rev.\ D {\bf 69}, 044005 (2004)
  [arXiv:astro-ph/0307034].
  Phys.\ Rev.\ D {\bf 69}, 124015 (2004).


\bibitem{Dev} D.~Jain, A.~Dev and J.~S.~Alcaniz,
  Phys.\ Rev.\ D {\bf 66}, 083511 (2002)
  [arXiv:astro-ph/0206224].
J.~S.~Alcaniz, D.~Jain and A.~Dev,
  Phys.\ Rev.\ D {\bf 66}, 067301 (2002).

\bibitem{Spergel} M.~Ishak, A.~Upadhye and D.~N.~Spergel,
  Phys.\ Rev.\ D {\bf 74}, 043513 (2006)
  [arXiv:astro-ph/0507184].

\bibitem{Linder} E.~V.~Linder,
  Phys.\ Rev.\ D {\bf 72}, 043529 (2005)
  [arXiv:astro-ph/0507263].


\bibitem{Tyson} L.~Knox, Y.~S.~Song and J.~A.~Tyson,
  %
  arXiv:astro-ph/0503644.


\bibitem{Maa} R.~Maartens and E.~Majerotto,
  Phys.\ Rev.\ D {\bf 74}, 023004 (2006)
  [arXiv:astro-ph/0603353].


\bibitem{Maart} K.~Koyama and R.~Maartens,
  JCAP {\bf 0601}, 016 (2006)
  [arXiv:astro-ph/0511634].


\bibitem{Carroll} I.~Sawicki and S.~M.~Carroll,
  arXiv:astro-ph/0510364.


\bibitem{Hu} Y.~S.~Song, I.~Sawicki and W.~Hu,
  arXiv:astro-ph/0606286.

\bibitem{DG} G.~R.~Dvali and G.~Gabadadze,
  Phys.\ Rev.\ D {\bf 63}, 065007 (2001)
  [arXiv:hep-th/0008054].


\bibitem{GS} G.~Gabadadze and M.~Shifman,
  Phys.\ Rev.\ D {\bf 69}, 124032 (2004)
  [arXiv:hep-th/0312289].

\bibitem{Ignat}I.~Antoniadis, R.~Minasian and P.~Vanhove,
  Nucl.\ Phys.\ B {\bf 648}, 69 (2003)
  [arXiv:hep-th/0209030].

\bibitem{Pierre} E.~Kohlprath,
  Nucl.\ Phys.\ B {\bf 697}, 243 (2004)
  [arXiv:hep-th/0311251].
E.~Kohlprath and P.~Vanhove,
  arXiv:hep-th/0409197.

\end{thebibliography}
\end{document}